\documentclass[12pt]{article}

\usepackage[affil-it]{authblk}
\usepackage[margin=1in]{geometry}
\usepackage{amsmath}
\usepackage{graphicx}
\usepackage{multicol}
\usepackage{textcomp}
\usepackage{listings}
\usepackage[dvipsnames]{xcolor}

\usepackage{titlesec}
\usepackage{amssymb}
\usepackage{graphicx}
\usepackage{caption}
\usepackage{float}
\usepackage{etoolbox}
\usepackage{subfig}

\usepackage{algorithm}
\usepackage{algpseudocode}

\usepackage[colorlinks=true,linkcolor=black,citecolor=blue,filecolor=magenta,urlcolor=cyan]{hyperref}

\begin{document}
\title{Effect of Demagnetization Method on Remnance Magnetization States in Metallic Ferromagnets}
\author{Jennifer S. Freedberg and E. Dan Dahlberg}
\maketitle

\begin{abstract}

Parametric plots of the remagnetization versus demagnetization remnances were found for four metallic ferromagnets -- nickel wire, two types of AlNiCo, and samarium cobalt 2:17. These plots, known as Henkel plots, were compared to Wohlfarth's model for noninteracting magnetic particles and several Preisach models. The remagnetization data were taken with a variety of paths to the net zero magnetization state. The resulting Henkel plots exhibit similarities to independent Monte Carlo simulations.  The differences can be mostly explained by considering that the magnetization in the metallic ferromagnets occur by domain wall motion.
\end{abstract}

\section{Introduction}
Parametric plots of the remagnetization remnance versus demagnetization remnance, commonly known as Henkel plots, previously have been used to explore interactions in a variety of magnetic systems \cite{Garca-Otero2000, Basso1994, Harres2013, Vajda1994}. They are typically compared to Wohlfarth's prediction \cite{Wohlfarth1958} for DC, AC, and thermally demagnetized samples based on noninteracting particles with uniaxial anisotropy at zero temperature. Prior research found that the Henkel plots for particulate media depended upon the zero magnetization state starting point for the remagetization remnance \cite{Bissell1989}. Following this, researchers developed several Preisach models including a classical Preisach model (PM),  moving Priesach model (MPM), and complete moving hysteresis model (CMH) to include the effects of the initial zero magnetization state \cite{Garca-Otero2000, Harres2013,Vajda1994,Bissell1989, Harrell1993}. 

Most of this research has been focused on particulate magnetic systems ignoring metallic systems. To explore the similarities and differences between particulate and metallic systems, we have undertaken an investigation of the metallic ferromagnets AlNiCo 2, AlNiCo 5, samarium cobalt 26 (SmCo) 2:17, and nickel. 

We used three distinct paths to obtain the required zero magnetization state for AlNiCo 2 and AlNiCo 5, and four for SmCo and nickel.  One of the four paths to net zero magnetization was AC demagnetization, created by applying a saturating AC field that decreases in magnitude with time. This process is expected to behave like thermal demagnetization as the number of AC oscillations approaches infinity \cite{Harres2013}. Another net zero magnetization state was DC demagnetization. This state is created by first applying a saturating field and then reversing the field direction until zero magnetization is reached. With this DC demagnetization state as the starting state, two different remagnetization paths were explored: applying the original saturating field either aligned or antialigned with the defined positive direction. Finally, for the SmCo and Ni, we used thermal demagnetization by raising the sample above its Curie temperature and then cooling in zero field to produce a statistically random state.

We compare our results with the various models developed to predict Henkel plots of particulate systems. We find that our metallic ferromagnets exhibit significant similarities to Wohlfarth's simple model and the Preisach models.  In what follows, we will discuss these models, our experimental procedures, and present our experimental results and discussion.

\subsection{Theory}
Wohlfarth developed a relationship between the demagnetization remanence and remagnetization remnance magnetizations in a set of noninteracting single domain particles, with uniaxial anisotropy \cite{Wohlfarth1958}. He predicted a simple relationship for a statistically random sample given by:

\begin{align}
\frac{I_D^{AC}(H)}{I_R(\infty)} = 1 -\frac{2I_R(H)}{I_R(\infty)},
\label{SW_AC}
\end{align}

\noindent where $I_D^{AC}(H)$ is the AC demagnetization remnance, $I_R(\infty)$ is the remagnetization remnance at saturation, and $I_R(H)$ is the remagnetization remnance. The normalization by $I_R(\infty)$ limits the demagnetization (remagnetization) remnance to be between -1 and 1 (0 and 1).

For the DC demagnetized state, the relationship between demagnetization remnances $I_D^{DC}$ and remagnetization remnances $I_R$ for a particulate system is also simple: 
\begin{align}
I_D^{DC}(H) = 
\begin{cases}
I_R(H_0)-2I_R(H) & H < H_0 \\
-I_R(H) & H > H_0,
\end{cases}
\label{SW_DC}
\end{align}

\noindent where $I_D$ is the demagnetization remnance, $I_R$ is the remagnetization remnance, and $H_0$ refers to the effective saturation (the latter occurs at a smaller magnitude of field than saturation). In his original paper, Wohlfarth only mentions one type of DC demagnetization. However, the original saturating field can be applied either aligned or antialigned with the defined positive direction; Wohlfarth did not discuss possible differences between these two and in our work we measured both remagnetization remanences. 

Because Wohlfarth did not take interactions into account, other models, such as the aforementioned PM, MPM, and CMH models were developed to better describe real systems with interactions. The Classical PM describes a system having a Gaussian distribution for both the critical field distribution and interaction field distribution \cite{Vajda1994}. The MPM describes a system with a moving parameter $\alpha$, which describes the effect the sample's magnetization has on itself \cite{doi:10.1002/mma.1670150302}. Finally, the CMH computes the reversible and irreversible magnetization components and their relationship to each other \cite{Vajda1994}.  Although Wohlfarth did not consider interactions, the simplicity of Wohlfarth's relations contribute to its continued popularity.

\section{Experimental Methods}
For the studies, fourteen samples were machined from stock, from there we obtained similar results within a sample and therefore only reported four for clarity. AlNiCo 2 and 5 were produced by a combination of milling with a carbide ball endmill and sanding to rectangular prisms of dimensions $0.25 \times 0.25 \times 1.27 cm^3$ and $0.127 \times 0.127 \times 1.02 cm^3$, respectively. The SmCo was produced by sanding to a rhomboid of approximate size $0.20 \times 0.254 \times 1.27 cm^3$ at its points. Nickel wire of length 1.67 with a diameter of 0.076 $cm$ was used. 

The initial magnetization state for all measurements was set by applying a magnetic field sufficient to saturate the sample. In the case of the AlNiCo samples the field was 9 kOe, the SmCo saturation field was 17 kOe, and for Nickel the field was 700 Oe. The DC demagnetized state was prepared by applying a saturating field followed by a field antiparallel to the original direction of saturation. The antiparallel field magnitude was increased to slightly above the coercive field (a sample dependent value), and returned to zero field.  As mentioned, two DC remagnetization paths were explored. One path, we call DC forwards remagnetization, consisted of making remanence measurements with the applied field parallel to the original saturating field.  For the second path we perform DC backwards remagnetization, where the applied field is antiparallel to the original saturating field. The AC demagnetized state was obtained by applying an alternating field that was decreased in magnitude. Lastly, we performed a thermal demagnetization for the SmCo and Ni wire by heating each sample above the Curie temperature and then cooling in zero field; this state is thus expected to be statistically random. The SmCo was thermally demagnetized at a temperature of 960\textdegree C, which is above its Curie temperature of 825\textdegree C but below its sintering temperature of 1050\textdegree C \cite{Cullity2009, ArnoldMagnetics}. The nickel was thermally demagnetized at 400 \textdegree C, above its Curie temperature of 358 \textdegree C \cite{Cullity2009}.

\section{Results and Discussion}

\begin{figure*}[!t]
\centering
\subfloat[AlNiCo 5]{\includegraphics[width=2.5in]{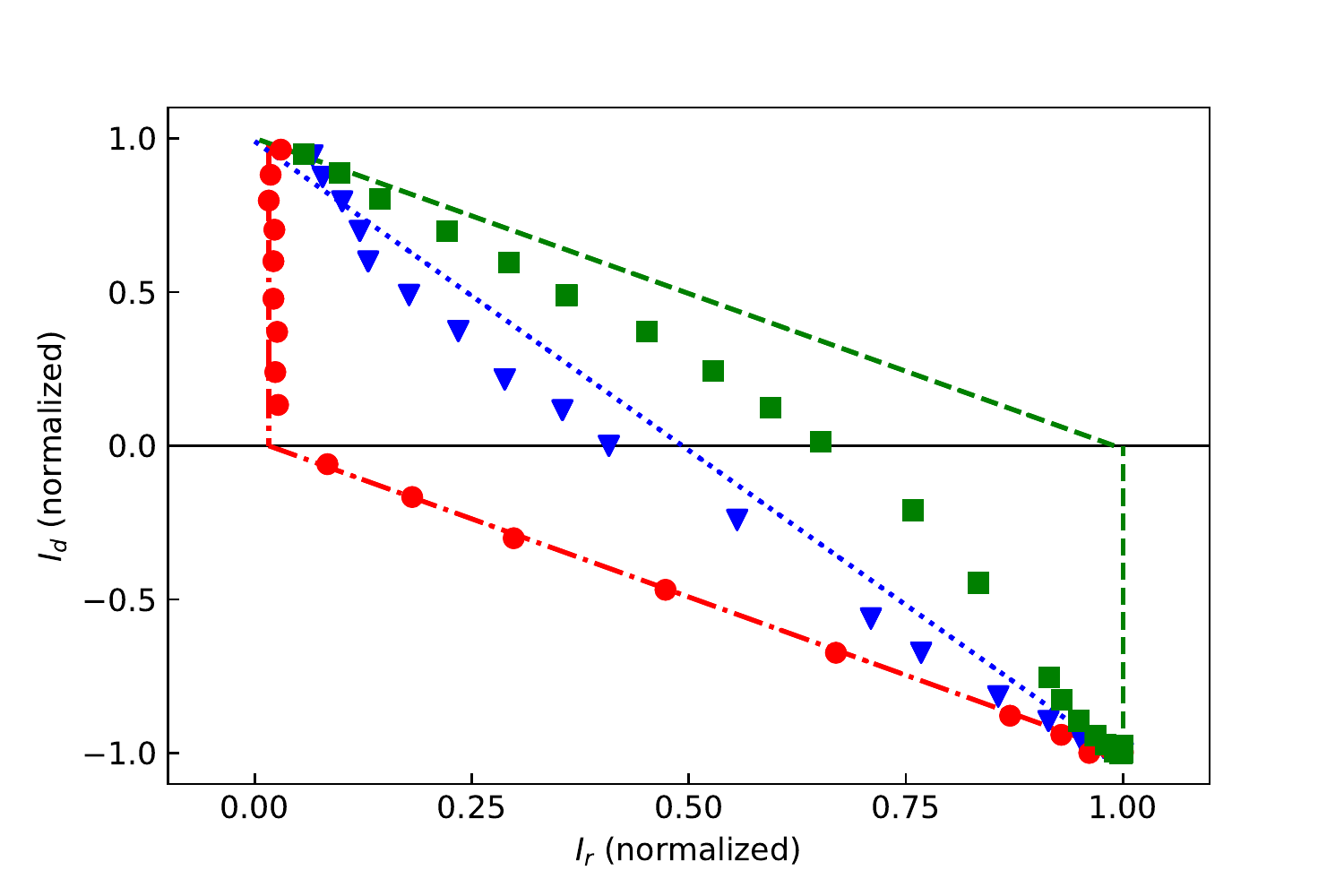}%
\label{A5}}
\hfil
\subfloat[AlNiCo 2]{\includegraphics[width=2.5in]{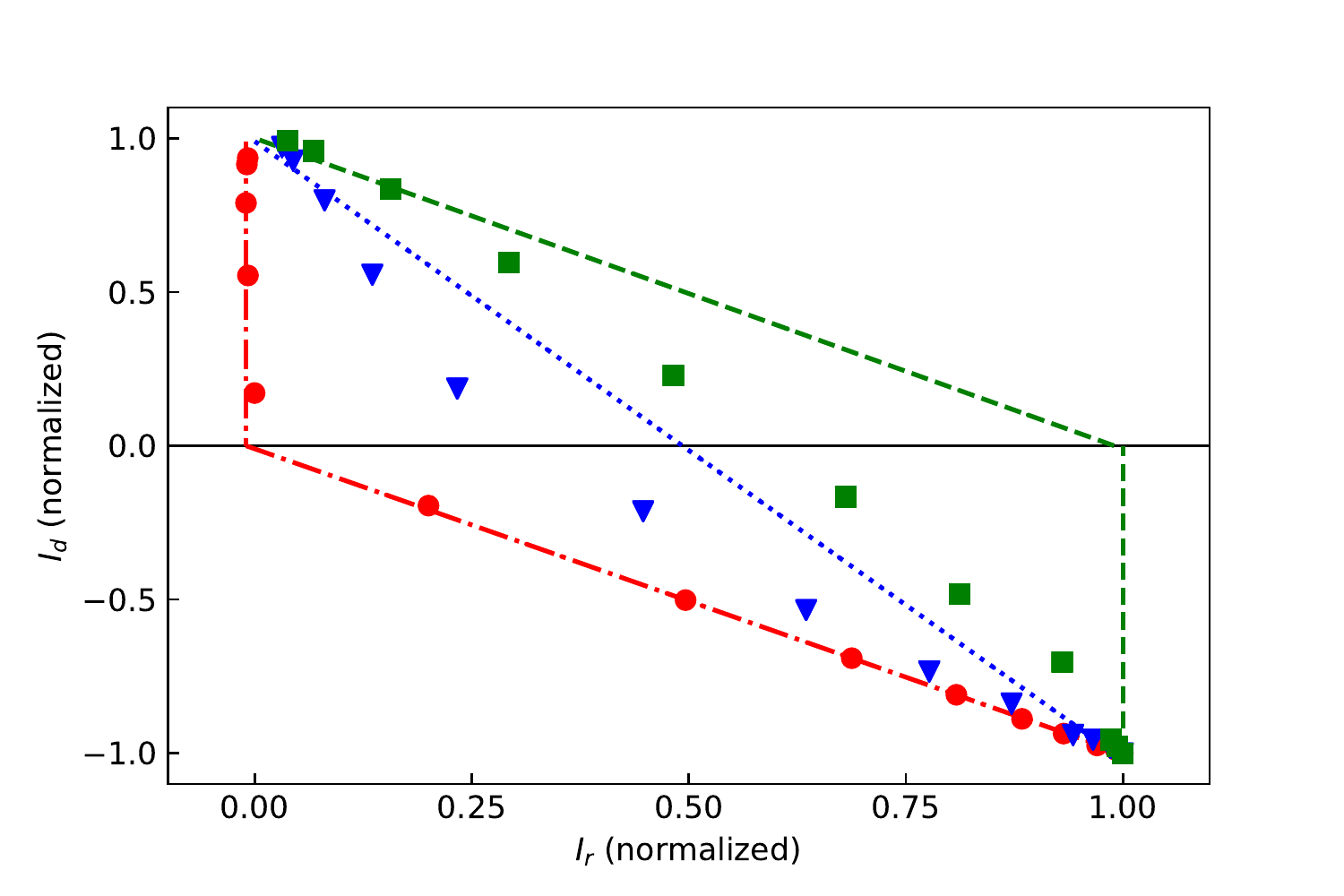}%
\label{A2}}\par
\subfloat[SmCo]{\includegraphics[width=2.5in]{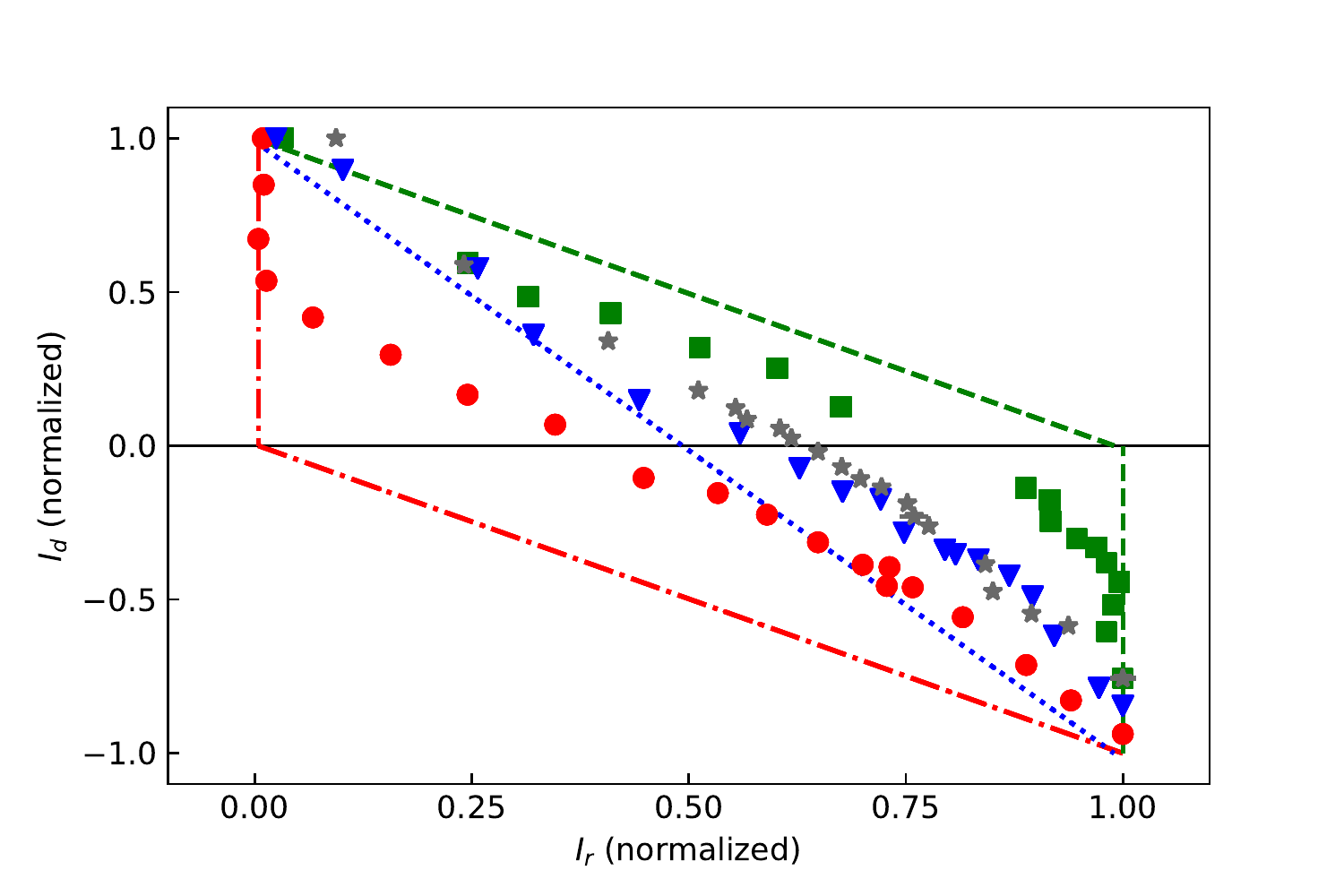}%
\label{SmCo}}
\hfil
\subfloat[Nickel Wire]{\includegraphics[width=2.5in]{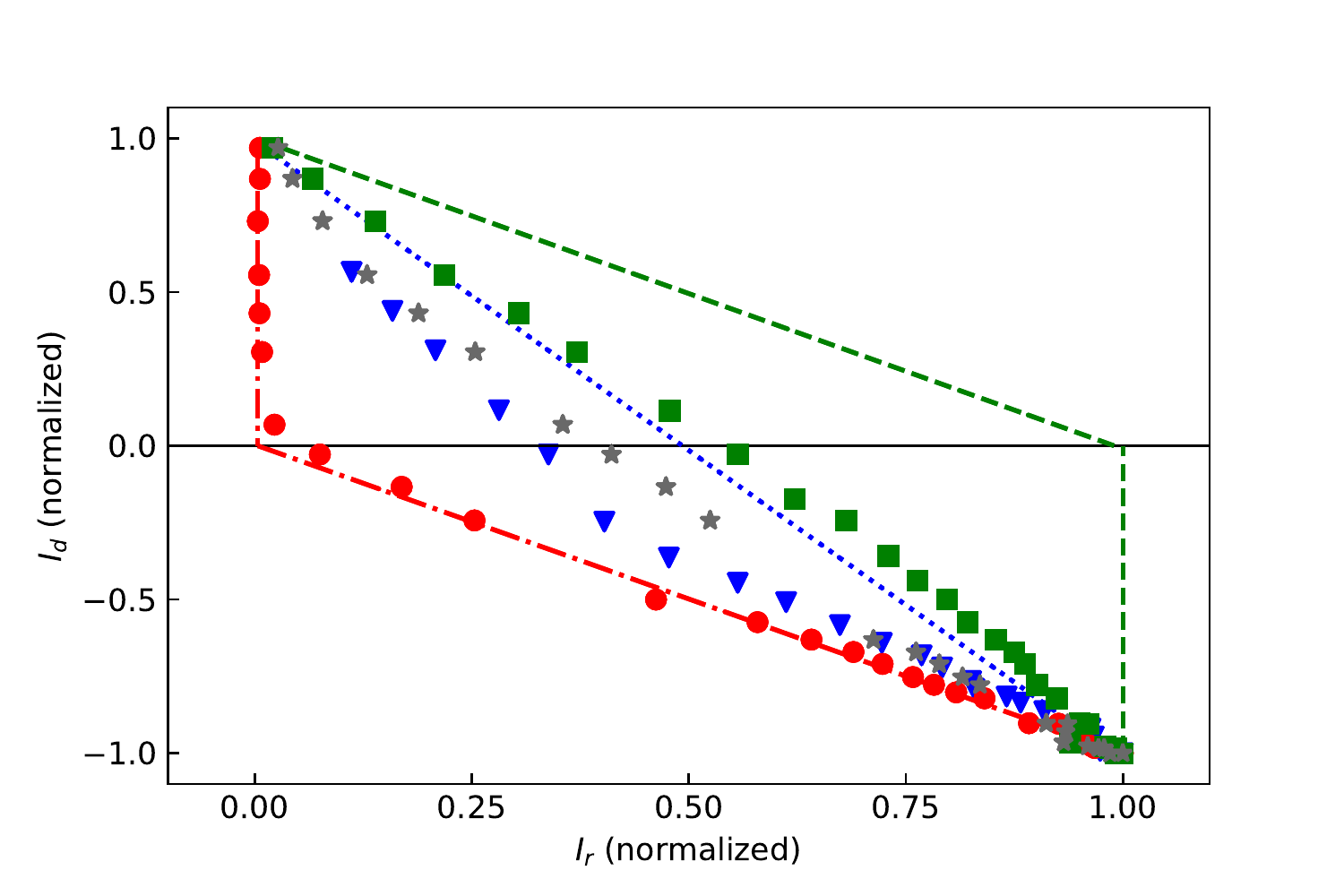}%
\label{Ni}}\par
\caption{\label{Full_Henkels} \textit{Henkel Plots for AlNiCo 5 (\ref{A5}), AlNiCo 2 (\ref{A2}), SmCo (\ref{SmCo}), and Nickel wire (\ref{Ni}). The y-axis is the normalized demagnetization remnance ($I_d$) and the x-axis is the normalized remagnetization remnance ($I_r$). The lines are the predicted Wohlfarth relations for reference. The red dot-dash line is DC backwards demagnetization, blue dotted line is AC demagnetization, and green dashed line is DC forwards demagnetization. Our experimental data for DC backwards demagnetization is a red circle, AC demagnetization is a blue triangle, and DC forwards demagnetization is a green square. Experimental thermal demagnetization data are grey stars. Error bars are plotted, but too small to be seen. (color online).}}
\label{fig_sim}
\end{figure*}

A summary of this study is presented in Figure \ref{Full_Henkels}. We first note that samples with demagnetization factors between ~0.05 to 0.5 were measured, and no significant difference between the data sets for a given sample was found. In what follows, we will discuss the Henkel plot for each type of demagnetization state after the following few general comments.  First, the magnetization process for metallic ferromagnets is different from particulate media. In the metallic ferromagnetic case, the magnetization processes occurs by the motion of domain walls and their pinning and depinning.  Second, we do not anticipate the ferromagnetic exchange interaction to play a major role since it is a short range interaction and the various regions are separated by domain walls.  This means one would not expect the Henkel plots to be effective at exhibiting positive or magnetizing exchange interactions. A model that considers only the exchange energy would result in Henkel plots consisting of mostly straight line segments.  Therefore the presence of anisotropy energies, the dipolar interactions between the domains, the distribution of wall pinning energies, and domain wall motion would therefore be responsible for the general shape the Henkel plots.  A last comment is that the same set of demagnetization remnance data is obtained independent of the path to zero magnetization. 

Now turning to the DC backwards data, since the zero magnetization state was produced by application of a field in this direction it is not surprising that there is no significant change in the remanance until fields larger than this field are applied.  This explains the abrupt initial drop in the DC backwards demagnetization curves in Figure \ref{Full_Henkels}.   The straight line behavior after the initial drop indicates a lack of sensitivity to dipolar interactions and domain wall pinning energy distributions.  This picture, however, does not address the behavior of the SmCo and Ni transition to a straight line at lower fields.  One possibility is there is an asymmetry in the pinning fields of the domain walls, i.e., the difference in the pinning energies is possibly due to a difference in the barrier heights of the pinning process. This effect is known as domain controlled nucleation, and is especially strong in SmCo \cite{Cullity2009}. 

The shape of the DC forwards demagnetized Henkel plots consistently deviated the most from the simple prediction. This is because in DC forwards demagnetization, domains with walls whose pinning energy is small are aligned against both the field and domains with walls of high pinning energy. In this case, the energy of the dipolar interactions and applied field become greater than the domain wall pinning energy present in the noninteracting case (just the Zeeman energy from the applied field). This suggests that domain walls are unpinned at lower applied fields than expected in the noninteracting limit.  We do not see this non-adherence in the DC backwards case because dipole interactions from domains whose walls with lower pinning energy produce weaker dipole fields, so this effect does not play a role in DC backwards adherence.


The AC and thermally demagnetized samples would be expected to only reflect dipolar interactions and wall pinning energy distributions.  As seen in the data, they are reasonably close to the straight line expected if there were no interactions.  The larger deviations from this are the AC demagnetized plots.  

In general the results in figures \ref{A5} and \ref{A2} display clear qualitative behavior for all four zero magnetization processes to what was seen in Monte Carlo simulations by using the Moving Preisach Model \cite{Vajda1994}. Specifically, these simulations show that DC backwards adheres well with the models, AC demagnetization curves downward from prediction and has a somewhat linear region in the middle, and DC forwards demagnetization curves down from the prediction with a slight bend towards the right hand side of the Henkel plot.

Monte Carlo simulations have also shown that isotropic materials or materials with cubic anisotropy deviate more from Wohlfarth's theory than materials with uniaxial anisotropy for AC and thermally demagnetized samples \cite{Garca-Otero2000}. The AlNiCo 5 (uniaxial anisotropy) is closer to the predicted Wohlfarth curve than AlNiCo 2 (isotropic) in states prepared using both AC and DC demagnetization. This behavior can be understood by recognizing that uniaxially anisotropic materials are forced to align either along or against the easy axis, and isotropic materials are not.

\section{Conclusion}
In each material, distinct Henkel curves were found for each type of demagnetization path as expected. Materials with stronger uniaxial anisotropy (AlNiCo 5) exhibited behavior closer to prediction than isotropic materials (AlNiCo 2). The Henkel plots most similar to Wohlfarth's model were those for the DC backwards and thermal demagnetization. 

One can assume that the interactions of both the applied field and harder domains created a higher effective field on the lower pinning energy wall the and therefore the DC Forwards demagnetization consistently fell well below the expected line. Additionally, the AC demagnetization curve was different from the thermal demagnetization for the SmCo and Ni from which one can conclude that AC demagnetization and thermal demagnetization will not produce the same Henkel plots.

\section*{Acknowledgment}

The authors would like to thank Austin Schleusner for help in the initial stages of the project and Kevin Booth's aide on several occassions.  We also thank Randall Victora for several productive conversations about Samarium Cobalt. This work was supported primarily by NSF GRANT DMR 1609782.\\ \\

\bibliographystyle{unsrt}
\bibliography{remnant_bib}

\begin{thebibliography}{10}

\bibitem{Garca-Otero2000}
J~Garcı{\'{a}}-Otero, M~Porto, and J~Rivas.
\newblock {Henkel Plots of Single-Domain Ferromagnetic Particles}.
\newblock {\em Journal of Applied Physics}, 87:7376, 2000.

\bibitem{Basso1994}
V.~Basso, M.~{Lo Bue}, A.~Magni, G.~Ummarino, and G.~Bertotti.
\newblock {Experimental study and theoretical interpretation of hysteresis
  loops and Henkel plots in soft magnetic materials}.
\newblock {\em Journal of Magnetism and Magnetic Materials}, 133(1-3):111--114,
  may 1994.

\bibitem{Harres2013}
A.~Harres, R.~Cichelero, L.~G. Pereira, J.~E. Schmidt, and J.~Geshev.
\newblock {Remanence plots technique extended to exchange bias systems}.
\newblock {\em Journal of Applied Physics}, 114(4):043902, jul 2013.

\bibitem{Vajda1994}
Ferenc Vajda, Edward {Della Torre}, and R~D McMichael.
\newblock {Demagnetized-state dependence of Henkel plots. I. The Preisach
  model}.
\newblock {\em Journal of Applied Physics}, 75:7376, 1994.

\bibitem{Wohlfarth1958}
E~P Wohlfarth.
\newblock {Relations between Different Modes of Acquisition of the Remanent
  Magnetization of Ferromagnetic Particles}.
\newblock {\em Journal of Applied Physics}, 29:7376, 1958.

\bibitem{Bissell1989}
P.R. Bissell, R.W. Chantrell, G.J. Tomka, J.E. Knowles, and M.P. Sharrock.
\newblock {Remanent magnetisation and demagnetisation measurements on
  particulate recording media}.
\newblock {\em IEEE Transactions on Magnetics}, 25(5):3650--3652, 1989.

\bibitem{Harrell1993}
J~W Harrell, David Richards, and Martin~R Parker.
\newblock {Delta-H plot evaluation of remanence behavior in barium ferrite
  tapes and disks}.
\newblock {\em Journal of Applied Physics}, 73:6722, 1993.

\bibitem{doi:10.1002/mma.1670150302}
Martin Brokate.
\newblock On the moving preisach model.
\newblock {\em Mathematical Methods in the Applied Sciences}, 15(3):145--157,
  1992.

\bibitem{Cullity2009}
B.~D. (Bernard~Dennis) Cullity and C.~D. (Chad~D.) Graham.
\newblock {\em {Introduction to magnetic materials}}.
\newblock IEEE/Wiley, 2009.

\bibitem{ArnoldMagnetics}
{Arnold Magnetics}.
\newblock {Samarium Cobalt Magnets | Arnold Magnetic Technologies}.

\end{thebibliography}
\end{document}